\documentclass[aps,preprint,showpacs,groupedaddress]{revtex4}
\usepackage[dvips]{graphicx}
\begin{document}

\title{Integrating fluctuations into distribution of resources in transportation networks}

\author{Shan He}
\author{Sheng Li}
\email{lisheng@sjtu.edu.cn}
\author{Hongru Ma}
\affiliation{Department Of Physics, Shanghai Jiao Tong University, Shanghai 200240, People's Republic of China}

\date{\today}

\begin{abstract}
We propose a resource distribution strategy to reduce the average travel time in a transportation network given a fixed generation rate. Suppose that there are essential resources to avoid congestion in the network as well as some extra resources. The strategy distributes the essential resources by the average loads on the vertices and integrates the fluctuations of the instantaneous loads into the distribution of the extra resources. The fluctuations are calculated with the assumption of unlimited resources, where the calculation is incorporated into the calculation of the average loads without adding to the time complexity. Simulation results show that the fluctuation-integrated strategy provides shorter average travel time than a previous distribution strategy while keeping similar robustness; the strategy is especially beneficial when the extra resources are scarce and the network is heterogeneous and lowly loaded.
\end{abstract}


\maketitle

\section{introduction}
Transportation networks, like many other complex networks, are the systems of interplay between structure and dynamics~\cite{ref1}, where the structure is the network topology and the dynamics is the transportation process, modelled as a large number of packets moving from source vertices to target vertices through edges. Examples of transportation networks are abundant in the real world, ranging from public transit networks, which transport passengers, to the Internet, which transports data. Different aspects of these networks have been covered by extensive researches in the literature, especially the robustness and the performance.

The robustness refers to the tolerance of a network to constituent damage without losing its characteristics. For a transportation network, the characteristics can be structural and dynamical, respectively. The structural characteristics are purely topological, including the existence of the giant component~\cite{ref3,ref4,ref5,ref6} and the average inverse path length~\cite{ref7,ref8}. It was found that scale-free networks, such as the World-Wide Web and the Internet~\cite{ref2}, display higher tolerance to errors but more vulnerability to attacks than random networks~\cite{ref3,ref4,ref5,ref6,ref7,ref8}. On the other hand, the dynamical characteristics are used in cases where the transportation load is redistributed upon the constituent damage. The load redistribution can trigger cascading failures~\cite{ref9,ref13}, such as the blackouts in the US's history~\cite{ref10,ref11,ref12}. It was made clear that cascading failures are only expected in heterogeneous networks when highly loaded vertices are removed~\cite{ref13}, and the sizes of the cascades can be drastically reduced by intentional removals of vertices having small loads~\cite{ref14,ref16}.

The performance has been defined as the maximum load of packets a network can support in the free state~\cite{ref18,ref24,ref29,ref33}. The enhancement of the performance is a key issue in transportation networks. It was shown that the performance could be enhanced by dynamical strategies, such as designing routing strategies~\cite{ref19,ref20,ref21,ref22,ref23}, manipulating edge weights~\cite{ref27,ref28} and reallocating resources~\cite{ref30}. Through the interplay between structure and dynamics, the transportation performance could also be enhanced by structural strategies~\cite{ref24,ref25,ref26}.

In this paper, we study the robustness and the performance of transportation networks, where a different definition of the performance is used. In previous works, the resources, which determine the capacities of vertices to process packets, usually had a prearranged distribution~\cite{ref19,ref20,ref21,ref22,ref23,ref24,ref25,ref26,ref27,ref28,ref29}. The enhancing methods chose to deviate packets from hub vertices in order to counterbalance topological heterogeneity~\cite{ref19,ref20,ref21,ref22,ref23,ref27,ref28}. The maximum load was boosted at the expense of sacrificing the short distances provided by the hub vertices and suffering the longer travel times of the packets. This trade-off might not pay off in some time-critical circumstances. Moreover, the exact boost on the maximum load was hard to predict. It might not be clear whether an enhanced transportation network could meet a load requirement. Here we assume that a load is demanded to be supported by a network and the distribution of resources is to be arranged. The performance is not defined by loads but by how close the average packet travel time is to the average shortest path length. This definition of the performance is more suitable than the previous one in a time-critical situation where a load demand is explicit.

To achieve high performance in the network, we first route packets along the shortest paths and allocate resources according to the average loads of the vertices. The allocated resources are essential for the network to support the demanded load in the free state~\cite{ref24}. Though on the average no congestion happens, packets are delayed on the paths due to the fluctuations, e.g., if the incoming packets on a vertex at a time step are more than the average, the excess packets have to wait on the vertex until later time steps. We then try to alleviate the delays by allocating some extra resources to the network. Often, the extra resources are limited, i.e., there are not enough resources to support packet processing without delays. A problem naturally arises: How to distribute the extra resources so that the performance is as high as possible?

We propose a simple two-step strategy to tackle the problem. First, we calculate the fluctuations in the transportation process, pretending that the extra resources are unlimited. Then, we distribute the extra resources according to the calculated fluctuations. The strategy has to be two-step because we aim to utilize the fluctuations to guide the resource distribution, but they are unknown before the resources are distributed. Extensive simulations demonstrate that the fluctuation-integrated strategy provides higher performance for a network than a previous distribution strategy while keeping similar robustness; the strategy is especially beneficial when the extra resources are scarce and the network is heterogeneous and lowly loaded.

The paper is organized as follows: We define the model for the transportation process in Sec. II. Sec. III is for the first step of the fluctuation-integrated strategy. Sec. IV is for the second step of the strategy as well as simulation results of the performance. The results of the robustness are shown in Sec. V. Finally, we conclude the paper in Sec. VI.

\section{the transportation model}
We consider a simple transportation process on an undirected unweighted network with $N$ vertices. Every vertex is equipped with a first-in-first-out queue which can be infinitely long. At each time step, a vertex $i$ generates a packet with a fixed probability $\rho$ ($0<\rho\leq 1$), and adds the packet to the back of its queue. The target of the packet is randomly selected among the other $N-1$ vertices. Then the vertex $i$ processes packets at the front of its queue. At most $C_i$ packets can be processed, where $C_i$ represents the resources distributed to the vertex. For a packet being processed, if the target is just $i$, the packet is removed from the network; Otherwise, the packet is forwarded to the next vertex on the shortest path to its target. If there are multiple shortest paths, one path is selected randomly. The forwarded packet is directly added to the back of the new queue, i.e., packets are assumed to pass through every edge in one time step. The state of this transportation process at a time step is essentially the distribution of the queue lengths. We update all the queues in parallel.

In this model, the average load on a vertex $i$, $\langle L_i \rangle$, is~\cite{ref24}
\begin{equation}
\langle L_i \rangle =2\rho +\rho \frac{B_i}{N-1},
\end{equation}
where $B_i$ is the betweenness centrality of the vertex $i$~\cite{ref31,ref32}. The first term on the right side of the equation is for packets that are generated or removed at the vertex $i$, while the second term is for those passing through $i$. The probability $\rho$ in the equation can be used to identify a demanded load in a network, as given the topological quantities $B_i$ and $N$, the average loads in the network are determined by $\rho$. Suppose that we have a network with average loads determined. Depending on the distribution of $C_i$, the network may undergo a continuous phase transition from the free state to the congested state~\cite{ref18,ref24,ref29,ref33}. When $C_i <\langle L_i \rangle$ for some $i$, there are packets congested in the network and the average travel time, $\langle T\rangle$, diverges. To achieve high performance of the network, we require $C_i >\langle L_i \rangle$ for any $i$ and call the sum of the average loads, $\sum _i\langle L_i\rangle$, the essential part of the resources allocated to the network. The remaining part, $\sum _i (C_i-\langle L_i\rangle)$, is called the extra part. The essential resources are distributed exactly according to the average loads and congestion can be avoided. However, there are packets delayed on the paths due to the fluctuations of instantaneous loads. We assume that the delays cannot be eliminated, because in real situations, the extra resources are usually limited, i.e., there are not enough resources to satisfy $C_i >L _i^*$ for any $i$, where $L _i^*$ denotes the largest value of the instantaneous load on a vertex $i$. The goal is to distribute the extra resources so that the delays are as short as possible. We first pretend that we have unlimited resources and calculate the fluctuations. Then we use the calculated fluctuations to guide the distribution of the extra resources. The two steps are discussed in the next two sections, respectively.

\section{the fluctuations for unlimited resources}
The case of unlimited resources is defined as $C_i >L _i^*$ for any $i$. It is not necessary that $C_i \to \infty$. In this case, the fluctuations, i.e., the standard deviations, can be calculated in a simple manner because every packet in a network moves freely and there are no packet-packet interactions.

Denote $\delta _{si}$ as the number of vertices that are reachable via a vertex $i$ (excluding $i$), if one starts at a vertex $s$ and follows shortest paths. The probability $p _{si}$ of a packet generated at $s$ being the incoming packet of $i$ is
\begin{equation}
p _{si}=\rho \frac{1+\delta _{si}}{N-1},
\end{equation}
where $i$ is included in the numerator. With the probability $1- p_{si}$, no packet from $s$ is expected at $i$. The corresponding standard deviation $\sigma _{si}$ is
\begin{equation}
[\sigma _{si}]^2=p _{si}-[p _{si}]^2.
\end{equation}
As there are no interactions between packets, the standard deviation $\sigma _i$ of the instantaneous load on the vertex $i$ is
\begin{equation}\label{fsigmai}
[\sigma _i]^2=\rho(1- \rho)+\sum _{s\neq i}[\sigma _{si}]^2,
\end{equation}
where the first term on the right side of the equation is for packets that are generated at $i$, while the second term is for those generated on other vertices. We have set the time window lengths of the statistics to unit time steps.

The quantity $\delta _{si}$ is called the dependency of a vertex $s$ on a vertex $i$, which is a byproduct of calculating the betweenness centrality, i.e., $B_i=\sum _{s\neq i}\delta _{si}$~\cite{ref34}. The algorithm for the betweenness centrality can be extended to include the above calculation. We find that the extension does not add to the time complexity $O(NM)$~\cite{ref34}, where $M$ is the number of edges. Note that the fluctuations we obtain are internal to the transportation model, not externally imposed~\cite{ref17}.

Another byproduct of calculating the betweenness centrality is the number of shortest paths $n _{vt}$ between a vertex $v$ and a vertex $t$~\cite{ref34}, which helps implement a routing table to route packets along shortest paths. According to the model, the routing table should tell a packet the next vertex to hop to, and in the case of multiple shortest paths, one path should be able to be selected randomly. We let the routing table $R_{st}$ hold all the candidates to hop to for a packet currently at a vertex $s$ and moving to a vertex $t$. Each candidate $v$ is assigned the weight $n _{vt}$ and one candidate is randomly selected with the probability proportional to the weight. The selected candidate is the next hop satisfying the requirements of the model. Note that the selection is a completely local activity. We build the routing table by breadth-first-search, which is a part of the betweenness centrality algorithm~\cite{ref34}.

We validate Eq.~(\ref{fsigmai}) through extensive simulations. The networks used in the simulations are constructed according to the configuration model~\cite{ref35,ref36} and the total connectedness of the networks are guaranteed. For each network constructed, we calculate the average loads and the fluctuations, build the routing table, all within the time complexity of calculating the betweenness centrality. Then we simulate the transportation process and measure the fluctuations. Various degree distributions and loads are tested. For the sake of clarity, we only show results for two scale-free networks. One is for the high load $\rho =0.90$, the other is for the low load $\rho =0.05$. The degree distribution of the two networks is $P(k)\sim k^{2.5}$ ($k\geq 2$).

\begin{figure}
\centering
\includegraphics[width=86mm]{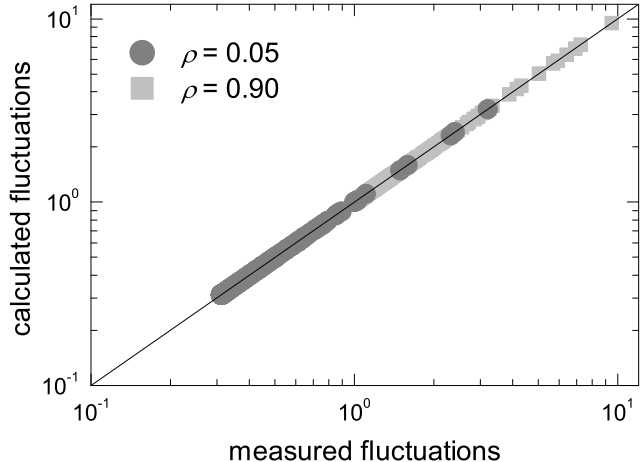}
\caption{Calculated fluctuations versus measured fluctuations for two scale-free networks in the simulations. The size of the two networks is $N=512$. One network carries the high load $\rho =0.90$, the other the low load $\rho =0.05$. The transportation process runs for $10^6$ time steps, where the first $10^5$ time steps are discarded. The solid line is a guide for the eye, which has the slope $1.0$ and passes through the origin.} \label{figure1}
\end{figure}

Figure~\ref{figure1} shows the comparison between the calculated fluctuations and the measured fluctuations for the two networks. As shown in the figure, the calculated fluctuations can be considered equal to the measured fluctuations. This is also true for the results we do not show. Therefore the calculation of the fluctuations is validated by the simulations.

\begin{figure}
\centering
\includegraphics[width=86mm]{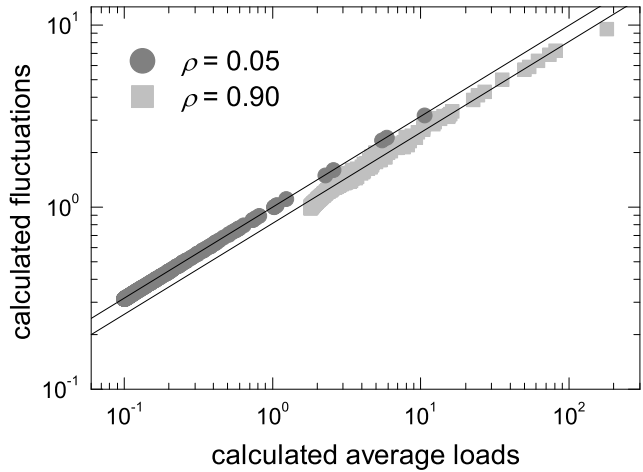}
\caption{Calculated fluctuations versus calculated average loads for the two networks shown in Fig.~\ref{figure1}. The two solid lines are guides for the eye. The slope of the lines is $0.5$.} \label{figure2}
\end{figure}

Recently, the relationship between fluctuations and average loads has attracted a lot of interests~\cite{ref37,ref38,ref39,ref40}. Since our calculation has been validated, we plot in Fig.~\ref{figure2} the calculated values of the average loads and the fluctuations in the two networks. The relationship can be described by a power law with the exponent $0.5$, though the power law is a little bit rough for the network of the high load. It has been shown that the exponent is always $0.5$ in the case of small average loads and small time window lengths, while in other cases the exponent is influenced by particular parameters~\cite{ref39}. We contribute that the fluctuations in the transportation model can be calculated for any average load.

\section{the resource distribution strategy}
After the fluctuations are calculated, we switch to the case of limited resources and propose the distribution strategy
\begin{equation}\label{fsigma}
C _i=\langle L _i\rangle +r\sigma _i,
\end{equation}
where the parameter $r$ is determined by the amount of the extra resources allocated to a network at a particular load, i.e., $r=\sum _i(C _i-\langle L _i\rangle)/\sum _i\sigma _i$. We refer to this strategy as the $\sigma$ strategy.

The $\sigma$ strategy is compared with a previous distribution strategy~\cite{ref13}
\begin{equation}\label{falpha}
C _i=\langle L _i\rangle +\alpha\langle L _i\rangle,
\end{equation}
which we refer to as the $\alpha$ strategy. Though the $\alpha$ strategy was originally proposed for studying cascading failures, Eq.~(\ref{falpha}) can be reinterpreted to fit the current context. The first term on the right side is for the essential resources, while the second term, which was called the tolerance term~\cite{ref13}, can be seen for the extra resources. The difference between the $\sigma$ strategy and the $\alpha$ strategy is only in the distribution of the extra resources. Thus the $\alpha$ strategy is more suitable for the comparison than other distribution strategies, such as constant~\cite{ref18,ref19,ref21,ref22,ref23,ref27,ref24,ref25,ref26}, proportional to degree~\cite{ref19,ref20,ref29} and more complicated strategies~\cite{ref28,ref41,ref42,ref43,ref44}. Empirical data have shown that real-world networks exhibit nonlinear resource distributions~\cite{ref45,ref46}. The extra resources distributed to vertices with smaller loads are relatively more. The nonlinearity can be understood as a trade-off between the cost of the resources and the robustness of the networks, where more weight is given to the cost for highly loaded vertices than lowly loaded ones~\cite{ref45}. The higher resource-to-load ratios of lowly loaded vertices might also be contributed by the physical limit on the minimum resources distributed to a vertex~\cite{ref46}, e.g., the bandwidth of a router has a lower limit, which bounds the minimum resources distributed to a site in the Internet even though the load of data can be much smaller. We will assume that no physical limit is imposed in the transportation model. And because both the $\sigma$ strategy and the $\alpha$ strategy are one-parameter, we do not consider the realistic case for simplicity.

The two strategies are compared by distributing the same amount of resources in a network. Given $\rho$ in the network, we use $r$ to represent the amount of the extra resources, and let $\alpha =r\sum _i\sigma _i/\sum _i\langle L _i\rangle$. The resources distributed to a vertex is a positive real number, whose integral part is the number of packets the vertex is able to process in a time step, while the decimal part is the probability of adding one more packet to the ability of the vertex in the time step. The performance of the resource distributions are evaluated by extensive simulations. Various values of $r$ are simulated given the high load $\rho=0.90$ and the low load $\rho=0.05$. For a pair of $r$ and $\rho$, we construct an ensemble of connected networks according to the configuration model~\cite{ref35,ref36} and run the transportation process. In each network, $10^4$ packets are sucessively sampled from the beginning of the transportation process. Then the first half of the packets are discarded. The remaining half are used to calculate the ensemble averaged travel time $\langle T\rangle$. We also obtain the ensemble averaged shortest path length $\langle D\rangle$ so that the performance can be evaluated by $\langle T\rangle - \langle D\rangle$.

\begin{figure}
\centering
\includegraphics[width=86mm]{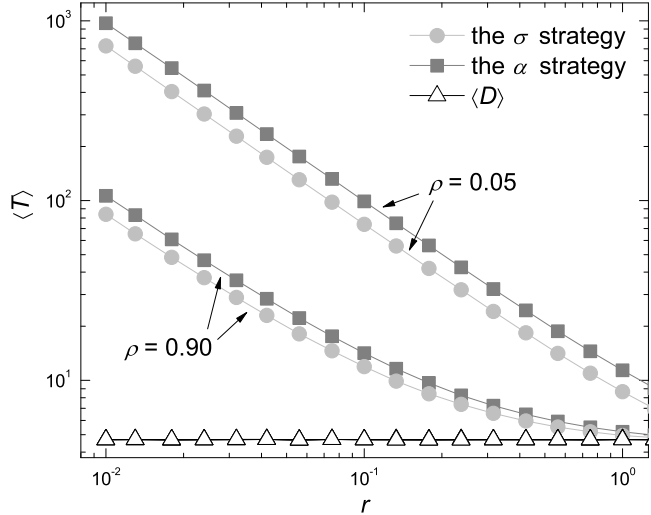}
\caption{The average travel time $\langle T\rangle$ versus the extra resources $r$ under the high load $\rho = 0.90$ and the low load $\rho=0.05$ for the $\sigma$ strategy and the $\alpha$ strategy. The average shortest path length $\langle D\rangle$ is shown as a reference. Each point in the figure is an average over $1000$ scale-free networks with degree distribution $P(k)\sim k^{2.5}$ ($k\geq 2$). The size of the networks is $N=512$.} \label{figure3}
\end{figure}

Figure~\ref{figure3} shows the results for scale-free networks. As shown in the figure, $\langle T\rangle$ of the $\sigma$ strategy is shorter than that of the $\alpha$ strategy while $\langle D\rangle$ is the same. Thus the performance of the $\sigma$ strategy is higher. It is also shown that the performance is influenced by both $r$ and $\rho$. When $r$ goes from small to large, $\langle T\rangle$ of both strategies approaches $\langle D\rangle$ and the differences between the two strategies become subtle (though still noticeable in the case of the low load). These trends are intuitive, as in the limit of $r \to \infty$, $\langle T\rangle$ is always $\langle D\rangle$ regardless of the strategies. Similar trends are observed when $\rho$ switches from low to high that $\langle T\rangle$ becomes shorter, and the differences between the two strategies turn out to be less significant. In the case of the high load, there are a large number of packet incoming events. The fluctuations relative to average loads are smaller than those in the case of the low load. As a result, the delays of packets are less probable and $\langle T\rangle$ is shorter. Moreover, The smaller relative fluctuations indicate the greater importance of the essential resources than the extra resources in Eq.~(\ref{fsigma}) and Eq.~(\ref{falpha}), i.e., higher loads correspond to smaller differences between the distribution of the extra resources. From the trends we can see that the high performance of the $\sigma$ strategy is especially beneficial when both $r$ and $\rho$ are small.

For more homogeneous networks like random networks, the loads and fluctuations are homogeneous. There is little difference whether the extra resources are distributed proportional to loads or fluctuations. We find that the performance of the $\sigma$ strategy is only a little higher than the $\alpha$ strategy (We do not show the data here.).

\section{dynamical robustness}
An important concern in transportation networks is the robustness. Originally, the $\alpha$ strategy distributed the extra resources to provide tolerance to cascading failures~\cite{ref13}. While the $\sigma$ strategy improves the performance by altering the distribution, it is a question whether the tolerance is preserved. To answer the question, we compare the dynamical robustness of the two strategies in the model of cascading failures~\cite{ref13}.

A cascading failure in a transportation network can be seen as having two stages~\cite{ref13,ref14}: The first stage is the initial attack, whereby the vertex with the largest average load is removed. The initial attack changes the distributions of the shortest paths and the average loads in the network. Though before the attack the resources are sufficient for every vertex according to either of the strategies, overloaded failures can occur after the attack. The second stage is the cascade of overloaded failures. Upon a distribution change, all the overloaded vertices are removed simultaneously. This causes a new change in the distribution. Subsequent failures occur until a change does not cause any overload. The damage of the cascade is quantified by the relative size $G=S'/S$, where $S$ and $S'$ are the sizes of the largest connected components in the network before and after the cascade, respectively. Small $G$ indicates severe damage. To reduce the damage of a cascade, an effective measure of defense can be implemented when the first stage of the cascade is encountered but the second stage is not~\cite{ref14}. The defense measure consists of intentional removals of a suitable fraction~\cite{ref16} of vertices that are initially lowly loaded.

\begin{figure}
\centering
\includegraphics[width=86mm]{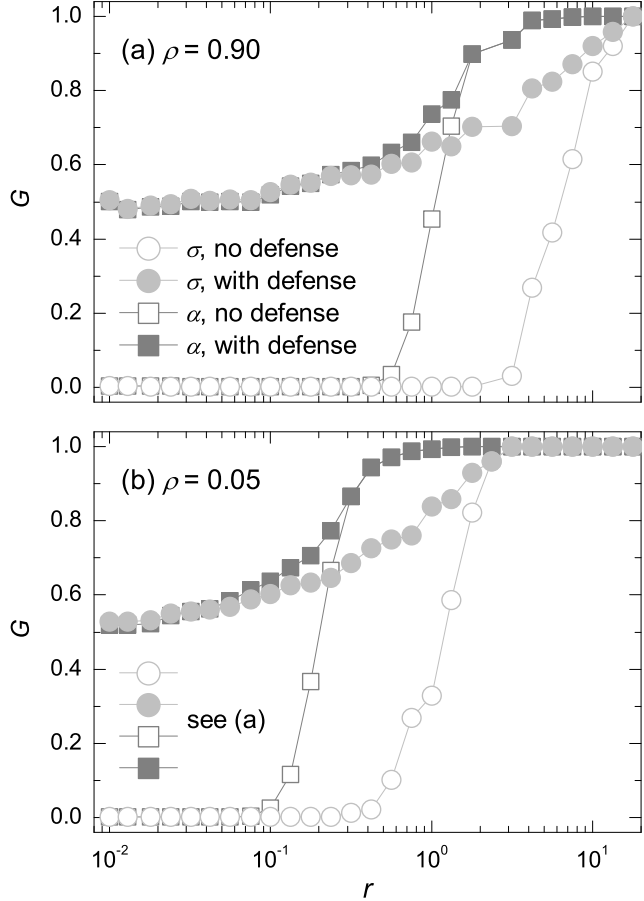}
\caption{The relative size $G$ versus the extra resources $r$ for scale-free networks under (a) the high load $\rho = 0.90$ and (b) the low load $\rho =0.05$. The figure compares the $\sigma$ strategy and the $\alpha$ strategy in or without the presence of the defense measure. Each point in the figure is an average over $100$ scale-free networks with degree distribution $P(k)\sim k^{2.5}$ ($k\geq 2$). The size of the networks is $N=5000$.} \label{figure4}
\end{figure}

Since cascading failures do not happen in random networks, we compare the $\sigma$ strategy and the $\alpha$ strategy only in scale-free networks. The networks are also constructed according to the configuration model~\cite{ref35,ref36}, but the total connectedness is not required. Figure~\ref{figure4}(a) shows the results as $G$ versus $r$ under the high load $\rho=0.90$. When $r$ is smaller than $0.4$, the two strategies are very similar in the robustness. For either strategy, $G$ is close to $0$ with the defense turned off while $G$ is a little larger than $0.5$ with the defense turned on. The similarity for small $r$ is due to the small difference between the distribution of the resources. When $r$ is larger than $10.0$, though the difference between the distribution is large, the two strategies are again similar in the robustness. The similarity coincides with the trend that $G \to 1$ as $r \to \infty$ regardless the strategies. For an intermediate $r$ between $0.4$ and $10.0$, $G$ of the $\alpha$ strategy is larger than that of the $\sigma$ strategy with the defense turned off. This is because fluctuations are usually sublinear functions of average loads~\cite{ref37,ref38,ref39,ref40}, the $\sigma$ strategy distributes more resources than the $\alpha$ strategy to lowly loaded vertices. These vertices are more robust under the $\sigma$ strategy, but they are responsible for the vulnerability to cascading failures and are the targets to remove in the measure of defense~\cite{ref14}. Although $G$ of the $\sigma$ strategy is greatly enlarged with the defense turned on, more resources are lost with the vertex removal in the defense. Thus we see the enhanced robustness of the $\sigma$ strategy is much closer to but still worse than that of the $\alpha$ strategy. Figure~\ref{figure4}(b) shows $G$ versus $r$ under the low load $\rho=0.05$. The results are generally the same except that the fluctuations are relatively larger; the $\alpha$ strategy is better than the $\sigma$ strategy between $r=0.1$ and $r=1.5$. Nonetheless, the dynamical robustness of the two strategies can be considered equal regardless of the loads when $0\leq r<0.1$. In this range of small $r$, the high performance of the $\sigma$ strategy is especially beneficial.

\section{conclusion}
In this paper, we have calculated the fluctuations in the transportation model and proposed the $\sigma$ strategy, which integrates the fluctuations into the distribution of resources. The performance and the robustness of the $\sigma$ strategy have been tested in the extensive simulations. We have found that the $\sigma$ strategy provides higher performance than the $\alpha$ strategy while keeping similar robustness, and is beneficial especially in a heterogeneous network with a low load and scarce resources. The results can be applicable in real-world transportation networks. For example, it was reported that on average more than $94\%$ of the available bandwidth in the Internet router network remains unused~\cite{ref45}, but one does have slow and unsteady web surfing experience every so often. Our work may present some hints on the solutions to such problems.

\begin{acknowledgments}
This paper was supported by the National Science Foundation of China
under Grant No. 10334020.
\end{acknowledgments}

\end{document}